\documentclass[12pt]{article}
\usepackage{graphicx}
\usepackage{bm}

\newcommand{\eq}[1]{Eq.~(\ref{#1})}
\newcommand{\eqs}[2]{Eqs.~(\ref{#1}) and (\ref{#2})}

\def\be{\begin{equation}}
\def\ee{\end{equation}}
\def\ba{\begin{eqnarray}}
\def\ea{\end{eqnarray}}
\def\half{{1 \over 2}}
\def\Tr{\mbox{Tr}}

\def\ome{\omega}
\def\tome{\widetilde{\omega}}
\def\ta{\tilde{a}}
\def\tn{\tilde{n}}
\def\Ome{\Omega}
\def\la{\langle}
\def\ra{\rangle}
\def\dphi{\dot{\phi}}
\def\cH{{\cal H}}
\def\cS{{\cal S}}
\def\nave{\bar{n}}
\def\dn{\delta n}
\def\hC{\hat{C}}
\def\hK{\hat{K}}
\def\tK{\tilde{K}}
\def\phys{{\rm ph}}
\begin{document}
\title{\Large {\bf The first heat: production of entanglement entropy in the 
early universe}}
\author{Sergei Khlebnikov and Akhil Sheoran \\ 
{\normalsize {\it Department of Physics and Astronomy, Purdue University}}, \\
{\normalsize {\it West Lafayette, IN 47907, USA}}
}
\date{}
\maketitle
\begin{abstract}
Entanglement entropy (EE) of a spatial region quantifies correlations
between the region and its surroundings. For a free scalar in the adiabatic 
vacuum in de Sitter space the EE is known to remain low, scaling as
the surface area of the region. Here, we study the 
evolution of entanglement
after the universe transitions from de Sitter to flat space. We
concentrate on the case of a massless minimally
coupled scalar. We find numerically that, after the de Sitter stage ends,  
the EE and the R\'{e}nyi entropy 
rapidly grow and saturate at
values obeying the volume law. The final state of the subsystem (region)
is a partially thermalized state reminiscent of 
a generalized Gibbs ensemble.
We comment on application of our
results to the question of when and how cosmological perturbations decohere.
\end{abstract}
\setcounter{footnote}{1}
\section{Introduction}

Entanglement entropy (EE) offers a way to characterize thermalization in isolated 
quantum systems. At late times, for a generic nonlinear system, the EE of a subsystem 
is expected to 
approach the thermodynamic entropy but, as recent work has
shown, surprisingly large amounts of it 
can be produced already at the early stages, well before
the nonlinear effects have had time to set in. For example, 
in a lattice gas 
escaping from a small container into a larger one, the initial growth of the EE can 
be attributed to free streaming \cite{HBosons}. 
The growth of the EE after
a global quench in an integrable field theory has also been ascribed to free streaming,
in that case, of quasiparticles produced by the quench 
\cite{Calabrese&Cardy,Casini&al,Cotler&al}. 

The examples recounted above correspond to particular choices of the initial 
out-of-equilibrium state. 
In the present paper, we look at yet another type of the initial state---a
highly squeezed quantum state inherited from an
earlier stage of the evolution. In cosmology, perhaps the best known example of this
is the state of super-horizon
perturbations that have been amplified (or, rather, have frozen in) during inflation
\cite{Mukhanov&Chibisov,Hawking:1982,Starobinsky:1982,Guth&Pi}.
The analog of the post-quench dynamics in this case is the evolution of the perturbations
{\em after} inflation, including the time of the second horizon crossing, when a given 
mode of perturbation reenters the horizon. We will assume that these large-wavelength 
perturbations remain in the linear regime, so they can 
be described by a free field theory.

The simplest model of the scale factor of the universe
in this context is to connect, at some value
$\eta_0$ of the conformal time, a de Sitter expansion, with the scale
factor $a(\eta) = -1 / H\eta$, where $H$ is a constant Hubble parameter, 
to a static flat universe with the scale factor
$a(\eta_0)$. One motivation for considering this problem is to understand better the
``thermal'' properties of cosmological perturbations.
Earlier work \cite{Brandenberger&al,Prokopec,Gasperini&Giovannini,Kruczenski&al} 
has used 
various coarse-graining procedures to define the entropy of those, resulting in fairly large 
values---for a massless minimally coupled scalar in de Sitter, of order $H^3$ 
per unit physical volume.
These results have been questioned later \cite{Kiefer&al} as implying too much 
decoherence at the time of the second horizon crossing, compared to what is allowed
by observations. Computations of the EE,
on the other hand, do not rely on coarse graining. We therefore ask: can 
entanglement
of a region with its surroundings lead to a finite density of the EE, and how large can 
that density be?

At first glance, the answer to the first question
may seem an obvious no: a finite density
implies an entropy scaling as the volume of an entangling region, while it is known
that the EE of a free scalar in the adiabatic vacuum in
de Sitter only scales as the surface area 
\cite{Maldacena&Pimentel}. Suppose, however, that we think not of the EE in de Sitter
itself but rather the one obtaining in a post-quench fashion after the de Sitter stage
has ended.
Studies of quenches in integrable theories \cite{Calabrese&Cardy,Casini&al,Cotler&al}
have shown that the entropy after a quench
saturates at a value obeying the volume law, and we might expect the same for
the present case. We start by verifying that expectation, using a numerical method.

Numerical calculations of the EE in Gaussian field theories are by now entirely standard
(see, for example, Ref.~\cite{Cotler&al} and references therein).
The novel element in our case is the dependence of the EE on the degree of 
squeezing, as
represented by the parameter $\eta_0$ defined above. The quantity we ultimately want
is the entropy per degree of freedom (DOF) or, more precisely, per DOF that has
been amplified during inflation. This can be computed on a lattice without a need to take
the continuum limit. The reason is that the frequency of the lattice modes is bounded
from above, so at a sufficiently small $|\eta_0|$ all the modes are amplified. To 
estimate the entropy per amplified mode, we can simply divide the entropy of a region by the 
number of lattice sites in it.

Fortuitously, a massless scalar in an expanding universe with
{\em two} spatial dimensions
is isomorphic in the infrared to the Bogoliubov phonon in 
an atomic superfluid with a variable coupling 
between the atoms.
The latter system can then be considered as a ``quantum simulator'' for the former.
In particular, 
a comparison of experimental data for the superfluid to the results presented here
may allow one to isolate the effects of non-Gaussianity that were written into
the quantum state during the time when 
the coupling was large. The question of how much information can be encoded in such
non-Gaussianities is of interest for both the early universe and the physics of 
black holes. 
Because of this potential application of our results, we have chosen to tailor our
presentation to the two-dimensional case, with an
occasional summary of results for three dimensions.

In two spatial dimensions (2d), we compute the EE and the second R\'{e}nyi entropy directly
for rectangular regions of various sizes. In three dimensions (3d), we use the following
modification. We consider, as the entangling region, a right rectangular column maximally
extended along 
the third coordinate $z$ and compute the entropy of the modes
independent of $z$, that is, having
momentum $k_z = 0$ in that direction. The number of these, for a column of square cross 
section with side length $L$ (in lattice units), is $L^2$. The volume law corresponds to the
entropy per DOF being $L$-independent, i.e., the entropy of the column scaling as $L^2$.
We find that, in both dimensionalities, the entropies of the regions grow rapidly 
after the end of the de Sitter stage, until they saturate
at values obeying the volume law. The saturation entropies depend logarithmically on the
parameter $\eta_0$. We discuss the significance of these logarithms in 
Sec.~\ref{sec:thermal}.

For a more precise characterization of the final state, we compute the distribution
of the thermal parameters $\beta_\ell$ \cite{Simon&al} corresponding  
to the Williamson normal modes. These are the modes for which the density matrix of the
subsystem (region) is diagonal. 
Those parameters can be compared to an {\em a priori} different set of 
thermal parameters, $f_\ell$, which determine the occupation numbers 
\be
\tn_\ell = (e^{f_\ell} - 1)^{-1} 
\label{tnell}
\ee
of the Bogoliubov transformed operators corresponding to the subsystem's usual
normal modes. $f_\ell$ can be computed directly in de Sitter under the assumption that
the subsystem is (quasi)isolated. They can be written as 
$f_\ell = f(\tome_\ell, \eta_0)$ where $f$ is a known function of the
normal frequency $\tome_\ell$ and the degree of squeezing. 
We find that, at large squeezing, 
the distributions of
$\beta_\ell$ (in the final state) and $f_\ell$ match fairly well. This suggests that the system 
undergoes a restricted version
of thermalization: at large enough times, the 
Williamson normal modes of a subsystem and its usual normal modes become more or less
the same.

The matching of the distributions of $\beta_\ell$ and $f_\ell$ lends credence to one 
of the coarse-graining proposal of 
Refs.~\cite{Brandenberger&al,Prokopec,Kruczenski&al}, namely, coarse-graining
in the occupation
number basis, provided that it is applied not in de Sitter itself but
rather after that stage has ended and the modes have been already 
oscillating for some time. The distributions do not match at earlier 
times. In addition, we recall that, unlike for coarse-graining,
the density matrix in our case is that of
a subsystem; the system as a whole remains in a pure state.

Matching of the distributions is also 
reminiscent of relaxation towards a generalized Gibbs ensemble (GGE)
in integrable models \cite{Rigol&al,Calabrese&al}. 
The difference is that, in our case, the particle numbers
$N_\ell$, to which the expectation values (\ref{tnell}) correspond,
are not strictly conserved,
because the subsystem is in fact not isolated. Our results are in keeping
with the general principle of statistical physics, according to which the 
equilibrium density
matrix of a subsystem should depend only on additive quasi-conserved
quantities, i.e., those that would be conserved if it were not for exchange of them
with the surroundings.

\section{The lattice model} \label{sec:lat}
The main object of our study is a Gaussian scalar field $\phi_i$ defined on 
a two-dimensional (2d) square lattice (although we also discuss the 3d case).
We consider two ways of defining dynamics for $\phi_i$. The first
is by discretizing the action of a continuum relativistic scalar minimally coupled
to gravity in a spatially flat expanding universe. The resulting action is
\be
S = \int d t  \left( \half \ta^2(t) \sum_i \dphi_i^2 - 
\half \sum_{ij} \phi_i \tK_{ij}(t) \phi_j \right) ,
\label{S}
\ee
where $\ta(t)$ is the scale factor of the universe,
\be
\tK_{ij}(t) = K_{ij} + m^2 \ta^2(t) \delta_{ij} \, ,
\label{tK}
\ee
$K_{ij}$ is a time-independent symmetric positive semidefinite matrix, and $m$ is
the mass of the scalar. Note that we have absorbed the area of the unit cell into 
$\ta^2$, so $\ta$ has dimension of time. We will present numerical results only for
the massless case, when $\tK_{ij} = K_{ij}$, but some of the intermediate expressions 
apply more generally.

The second approach to defining dynamics is to start with the 2d Bose-Hubbard
model in the superfluid phase,\footnote{For the phase diagram of the Bose-Hubbard model, 
see Ref.~\cite{Fisher&al}.} where the scalar is the Bogoliubov phonon.
On a square lattice, the Bose-Hubbard Hamiltonian is
\be
\cH_{BH} = \sum_i \left[ -J \sum_{e} (a_i^\dagger a_{i+e} + a_{i+e}^\dagger a_i) 
+ \half U n_i^2 - (\mu + \half U ) n_i \right] ,
\label{BH}
\ee
where $e$ goes over the positive lattice directions, 
$n_i = a_i^\dagger a_i$, and $J, U, \mu$ are parameters. We take $J > 0$ and $U > 0$, the
latter corresponding
to a repulsive interaction. In the superfluid phase, at low
energies, the number density $n_i$ is close to its average $\nave$, i.e.,
\be
n_i = \nave + \dn_i \, 
\label{dn}
\ee
where $\dn_i$ is small. Defining the order parameter phase $\theta_i$ via 
$a_i = \sqrt{n_i} e^{-i\theta}$ and expanding in small $\dn_i$ and $\theta_{i+e} -\theta_i$,
one obtains the quadratic Hamiltonian
\be
\cH_{BH,2} = J \nave \sum_{i,e} \left[ (\theta_{i+e} -\theta_i)^2 
+ \frac{1}{4 \nave^2} (\dn_{i+e} - \dn_i)^2 \right] + \half U \sum_i \dn_i^2  \, .
\label{BHquad}
\ee
Define symmetric matrix $L_{ij}$ such that
\be
\sum_{ij} \theta_i L_{ij} \theta_j = \sum_{i,e} (\theta_i - \theta_{i+e})^2 \, .
\label{Lij}
\ee
To identify the Bogoliubov phonon, one rescales $\theta_i$ and $\dn_i$ into
a new canonical pair $\phi_i$, $\pi_i$, as follows: 
$\phi = \hat{Z}^{-1/2} \theta$, $\pi = \hat{Z}^{1/2} \dn$ (in matrix notation), where
$\hat{Z}$ is made of the following matrix elements:
\be
Z_{ij} = \frac{1}{2 J \nave} \left( \delta_{ij} + \frac{J}{2 \nave U} L_{ij} \right) \, .
\label{Zij}
\ee
Suppose for a moment that the parameters in (\ref{Zij}) are time-independent.
Then, the new quadratic Hamiltonian is simply \eq{BHquad} rewritten in 
the new variables:
\be
\cH_{BH,2} = \half \phi \left( \hat{L} + \frac{J}{2 \nave U} \hat{L}^2 \right) \phi
+ J \nave U \pi^2 .
\label{BHquad2}
\ee
To establish a connection with the cosmological model (\ref{S}), we consider the case
when $\nave$ is large enough for the second term in the brackets in (\ref{Zij})
and (\ref{BHquad2}) to
be always negligible. If we neglect this term and switch time-dependence on in $U$ 
but not in
$J$ or $\nave$, the matrix (\ref{Zij}) will remain time-independent, and the expression
(\ref{BHquad2}) will still apply. The theory then becomes equivalent to one with the action
(\ref{S}) with zero mass and the following identifications: $\hat{K} = \hat{L}$ and
$\ta^2(t) = [2 J \nave U(t)]^{-1}$.

For a possible experimental realization, it is important that a simulation of the 
theory (\ref{S}) 
with atoms in an optical lattice requires only time-dependence of $U$ and not, for
instance, of the hopping matrix element $J$. This is specific to two spatial dimensions.

The sum $\delta N = \sum_i \dn_i$ represents a
variation of the total particle number and, if the latter is conserved, must be zero.
The Hamiltonian (\ref{BHquad}) guarantees that, if imposed initially, this condition
will hold at all times:
\be
\frac{d}{dt} \sum_i \dn_i = - 2 J \nave \sum_{ij} L_{ij} \theta_j = 0 \, .
\label{cons}
\ee
This is because $\sum_i L_{ij} = 0$, meaning that the symmetric matrix $\hat{L}$ has
a constant as its zero mode.
In the continuum limit, the particle number is conserved by the Neumann 
($\nabla \theta = 0$) boundary
condition on $\theta$ but not by the Dirichlet (fixed $\theta$) 
one. Eq.~(\ref{S}) with $\hK = \hat{L}$ can be viewed as a discretization of a
continuum action with the field subject to the Neumann condition.

In what follows, we remove the constant mode from the consideration entirely 
by assuming that we 
work with the ensemble with a fixed total particle number.

Upon transition to the conformal time $\eta$ via $d\eta = dt / \ta(t)$, the action 
becomes
\be
S = \int d\eta a(\eta) \left( \half \sum_i {\phi'_i}^2 
- \half \sum_{ij} \phi_i \tilde{K}_{ij} \phi_j \right) ,
\label{Sconf}
\ee
where $a(\eta) \equiv \ta[t(\eta)]$, 
and a prime denotes a derivative with respect to $\eta$. Recall that $\ta$ has dimension
of time, so $\eta$ is dimensionless.

The canonical momentum conjugate to $\phi_i$ is
\be
\pi_i = \ta^2(t) \dphi_i = a(\eta) \phi'_i 
\label{pi_i}
\ee
and can be obtained from either (\ref{S}) or (\ref{Sconf}).

We adopt the following form of the scale factor
\be
a(\eta) = \left\{ \begin{array}{ll} -1 / H \eta, & \eta \leq \eta_0  \, \\
-1 / H \eta_0 , & \eta > \eta_0 \, , \end{array} \right. 
\label{aeta}
\ee
corresponding to a de Sitter expansion that connects, at some
$\eta = \eta_0 < 0$, to a flat spacetime.

To obtain the mode expansion of the Gaussian field in this spacetime, we first diagonalize
the matrix $\hK$ by an orthogonal transformation
\be
\phi_i = \sum_k q_k O_{ki} \, ,
\label{orth}
\ee
resulting in a set of eigenvalues $\lambda_k = \ome_k^2$ (as noted above, we only
keep the nonzero ones) and the corresponding eigenvectors (the normal modes), 
which are the rows of the
orthogonal matrix $\hat{O}$; $q_k$ are the new canonical coordinates. 

The time-dependence of the amplitudes $q_k$ is obtained by solving a linear differential
equation. The result is as follows.
At $\eta < \eta_0$,
\be
q_k(\eta) = \half \sqrt{\pi H} \left[
b_k \eta H_\nu^{(1)}(-\ome_k \eta) + b_k^\dagger \eta H_\nu^{(2)}  (-\ome_k \eta) \right] ,
\label{qk}
\ee
where $b_k^\dagger$ are time-independent creation operators,
$H_\nu$ are the Hankel functions, and
\be
\nu^2 = 1 - \frac{m^2}{H^2} \, .
\label{nu}
\ee
The latter is for the general massive case (\ref{tK}). The normalization coefficient
in (\ref{qk}) is chosen so that $q_k$ and $p_k = a(\eta) q'_k$ are correct canonical pairs.

At $\eta > \eta_0$,
\be
q_k(\eta) =\frac{1}{\sqrt{2 \Ome_k a(\eta_0)}} \left(
c_k e^{-i\Ome_k \eta} + c_k^\dagger e^{i\Ome_k \eta} \right) ,
\label{qkflat}
\ee
where
\be
\Ome_k^2 = \omega_k^2 + m^2 a^2(\eta_0) \, ,
\label{Ome2}
\ee
and $c_k^\dagger$ is another set of time-independent operators. By matching the expansions 
(\ref{qk}) and (\ref{qkflat}) at $\eta = \eta_0$, one expresses one set of operators
in terms of the other.

At fixed $\nu$, the only dependence of \eqs{qk}{qkflat} on the Hubble
parameter
is in the prefactors: $q_k$ scales as $\sqrt{H}$ and its conjugate momentum $p_k$ as
$1/\sqrt{H}$. This dependence can be removed by a canonical transformation without 
affecting the value of the entropy. So, in computations of the entropy we can set
$H=1$. When we wish to express results in terms of the physical frequency
\be
\ome_{k,\phys} = \ome_k / a(\eta_0) \, ,
\label{ome_phys}
\ee
we will restore $H$. For example, the condition that a given mode is significantly 
amplified in de Sitter can be written as
\be
\ome_k |\eta_0| = \ome_{k,\phys} / H \ll 1 \, .
\label{ampl}
\ee
Another place where $H$ plays a role is the relation between the conformal time $\eta$ 
and the laboratory time $t$ during the flat space segment of the evolution:
\be
t = H \eta / |\eta_0| + \mbox{const.}
\label{ltime}
\ee
Note that, for a given $H$, the same interval of $\eta$ corresponds to different 
intervals of $t$ for different values of $\eta_0$.

\section{Entanglement entropy} \label{sec:ent}
At $\eta\to -\infty$, all modes (on a finite lattice, with the constant mode excluded) 
evolve adiabatically, and
the vacuum of the operators $b_k$ from (\ref{qk}) is the usual adiabatic vacuum of a 
free scalar in de Sitter. This is the state for which we will be studying
correlations as represented by the EE and the R\'{e}nyi entropy for different spatial
regions. The regions are defined by selecting a set of lattice sites and
are time independent.
We refer to these regions as subsystems.

To obtain the entropies, we first
construct the density matrix $\rho_A$ 
of a region $A$ by tracing out the local degrees of freedom
$(\phi_i, \pi_i)$ corresponding to lattice sites outside the region. 
The EE and the $q$-th R\'{e}nyi entropy of the subsystem are then defined by the usual 
formulas 
\ba
S_E & = & - \Tr (\rho_A \ln \rho_A) \, , \label{cSE} \\
S_R(q) & = & - \frac{1}{q - 1} \ln \Tr (\rho_A^q) \, . \label{cSR}
\ea
A convenient way to define $\rho_A$ for a Gaussian state is to require that it
reproduces the covariance matrix 
\be
C_{IJ}(\eta) = \half \la \{ \xi_I(\eta) , \xi_J(\eta) \} \ra ,
\label{CIJ}
\ee
where the braces denote anticommutator, and the vector 
\be
\xi = (\phi_1, \dots, \phi_n, \pi_1, \dots, \pi_n) ,
\label{xi}
\ee
whose components $\xi_I$ appear in (\ref{CIJ}),
combines the coordinates and momenta for the lattice sites {\em inside} the region $A$.
The averaging is over the adiabatic vacuum.
For a Gaussian state with zero $\la \phi_i \ra$  and $\la\pi_i \ra$, such as the one here, 
the covariance matrix carries all the information there is about the state.
The averages in (\ref{CIJ}) are then reinterpreted as those computed in a mixed state
in a Gaussian theory that contains only the variables in $A$, with the help
of a density matrix $\rho_A \equiv \rho_A(\eta)$.

The $2n \times 2n$ matrix $\hC$ 
that has $C_{IJ}$ as its matrix elements has the following normal
form, called the Williamson normal form \cite{Williamson}: 
$\hC = \hat{\cS}^T \mbox{diag}(\bm{\gamma},\bm{\gamma}) \hat{\cS}$,
where $\hat{\cS}$ is a symplectic matrix. The $n$ entries $\gamma_\ell$ that 
comprise
the vector $\bm{\gamma}$ are called the symplectic eigenvalues of $\hC$. The density
matrix $\rho_A(\eta)$ can be written as 
a product of thermal density matrices for a set of oscillators 
$a_\ell(\eta)$, $a_\ell^\dagger(\eta)$ corresponding to the Williamson normal 
modes, as follows \cite{Simon&al}:
\be
\rho_A(\eta) = \prod_{\ell = 1}^n (1 - e^{-\beta_\ell(\eta)}) e^{-\beta_\ell(\eta) 
a_\ell^\dagger(\eta) a_\ell(\eta)} \, ,
\label{rhoA}
\ee
where  
\be
\beta_\ell = \ln \frac{\gamma_{\ell} + 1/2}{\gamma_{\ell} - 1/2} \, .
\label{beta}
\ee
The entropies (\ref{cSE}) and (\ref{cSR})
are then the sums of those of the individual oscillators, that is, 
$S_E = \sum_\ell s_E(\gamma_\ell)$ and 
$S_R(q) =\sum_\ell s_R(q; \gamma_\ell)$, where
\ba
s_E(\gamma) & = & \gamma_+ \ln \gamma_+ - \gamma_- \ln \gamma_- \, , \label{SE} \\
s_R(q; \gamma) & = & \frac{1}{q - 1} \ln ( \gamma_+^q - \gamma_-^q ) \, , \label{SR} 
\ea
and $\gamma_\pm = \gamma \pm \half$. This method was used for a calculation of the
entropy growth after a quench in a Gaussian theory in flat spacetime in 
Ref.~\cite{Cotler&al}.

We now present numerical results for the model (\ref{Sconf}) on a square 2d lattice with
$K_{ij}$ set equal to $L_{ij}$, \eq{Lij}, and $a(\eta)$ of the form (\ref{aeta}).
The eigenvalues of this $\hK$, for an $M\times N$ lattice, are
\be
\ome_k^2 = 4 - 2 \cos k_x - 2 \cos k_y \, ,
\label{ome2}
\ee
where $k_x$ is quantized in units of $\pi / M$, and $k_y$ in units of $\pi / N$.
Note that $\ome_k$ is bounded from above by $2\sqrt{2}$, so for a sufficiently small
$|\eta_0|$ all lattice modes are amplified.

With the Hubble parameter set to one 
as described at the end of Sec.~\ref{sec:lat}, the only parameters
left in the model 
are the mass parameter $\nu$, \eq{nu}, and the transition time $\eta_0$
in \eq{aeta}. We can also vary the total size of the system and the size and 
shape of the entangling region. Here, we present results for the massless
case $\nu = 1$ and the total lattice size of $31\times 33$.

\begin{figure}
\begin{center}
\includegraphics[width=4in]{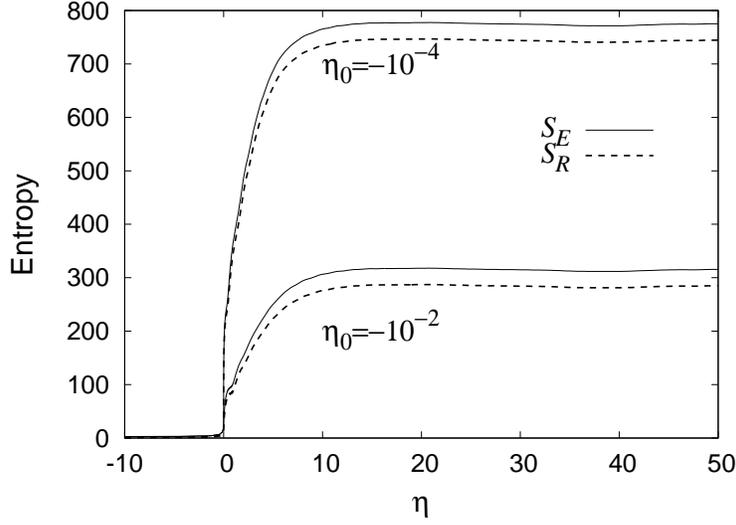}
\end{center}                                              
\caption{\small Time dependence of the EE (solid lines) and the second R\'{e}nyi
entropy (dashed lines) for a $10\times 10$ square in
the middle of a $31\times 33$ system. Larger entropies correspond to $\eta_0 = -10^{-4}$,
and smaller to $\eta_0= -10^{-2}$.}
\label{fig:tdep}  
\end{figure}

Fig.~\ref{fig:tdep} shows the EE and the second ($q=2$) R\'{e}nyi entropy 
for a $10\times 10$ square
in the middle of the system for two different values of $\eta_0$.
We see that, after the expansion is switched off, the entropies grow rapidly,
until they saturate at values that are large compared to the entropy the region had 
in de Sitter. 

\begin{figure}
\begin{center}
\includegraphics[width=4in]{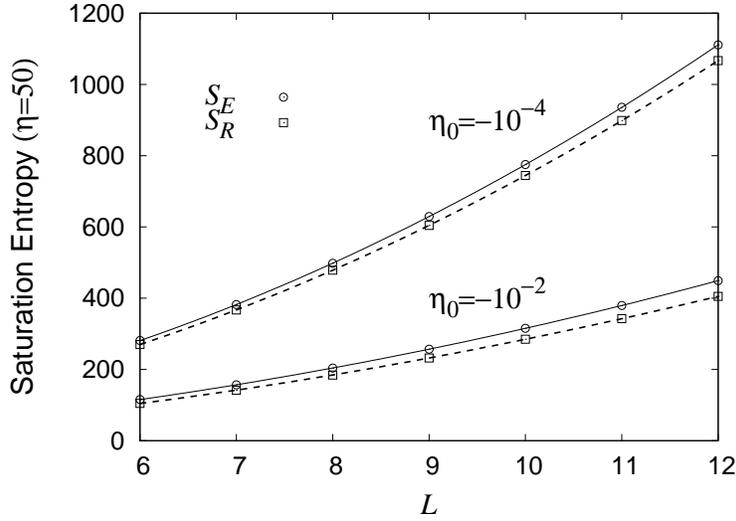}
\end{center}                                              
\caption{\small Dependence of the saturation entropies,
defined as those at $\eta = 50$, on $L$ (in lattice units)
for $L\times L$ squares in the middle of a $31\times 33$ system. Circles represent
the EE, and squares the second R\'{e}nyi entropy.
The lines are fits by quadratic polynomials $b L^2 + c L + d$.
Larger entropies correspond to $\eta_0 = -10^{-4}$, and smaller to $\eta_0= -10^{-2}$.}
\label{fig:squares}  
\end{figure}

The dependence of the saturation entropy on the subsystem linear size $L$ is shown in
Fig.~\ref{fig:squares}. 
One finds a good volume scaling with a coefficient of $L^2$ that depends
on $\eta_0$. That dependence, for $|\eta_0| \leq 0.01$, is well fit by 
\ba
b_E(\eta_0)[2{\rm d}] & = & - \ln |\eta_0| - 1.8 \, , \label{fitE} \\
b_R(\eta_0)[2{\rm d}] & = & - \ln |\eta_0| - 2.1  \label{fitR}
\ea
for the EE and the second R\'{e}nyi entropy, respectively. Because for these
small $|\eta_0|$ all modes we have on the lattice are amplified, we can view
(\ref{fitE}) and (\ref{fitR}) as estimates of the entropies per degree of freedom
(DOF) of the subsystem.

For a parallel computation in three dimensions, 
we consider an entangling region in the form of a right vertical 
column that extends for all values of the 
third ($z$) coordinate and has a constant cross-section of $L\times L$ sites in the
$(x,y)$ plane. To estimate the entropy per an amplified DOF, 
we restrict
attention to the $k_z = 0$ mode of the field. Then, the total number of DOF in the 
subsystem is $L^2$, just as in the 2d case. The difference with that case is in the form
of the mode functions at $\eta < \eta_0$: instead of (\ref{qk}) we now have 
\be
q_k(\eta) = \half \sqrt{\pi} H (-\eta)^{3/2} 
\left[ b_k H^{(1)}_\nu (-\ome_k \eta) + b_k^\dagger H^{(2)}_\nu (-\ome_k \eta) \right] ,
\label{qk_3d}
\ee
where
\be
\nu^2 = \frac{9}{4} - \frac{m^2}{H^2} \, .
\label{nu_3d}
\ee
The results are similar to those for two dimensions. For the massless case ($\nu = 3/2$),
the counterparts of \eqs{fitE}{fitR} are
\ba
b_E(\eta_0)[3{\rm d}] & = & - 2 \ln |\eta_0| - 1.9 \, , \label{fitE_3d} \\
b_R(\eta_0)[3{\rm d}] & = & - 2 \ln |\eta_0| - 2.2  \label{fitR_3d} 
\ea
for $|\eta_0| \leq 0.01$.

Finally, we record an estimate for the entropies per unit physical volume, corresponding 
to the saturation values (\ref{fitE_3d}) and (\ref{fitR_3d}). In terms of the
physical frequency (\ref{ome_phys}), the logarithms in (\ref{fitE_3d})--(\ref{fitR_3d})
become those of $\ome_{k,\phys} / H$, so
\be
S_E/V_{\phys} \sim \int \frac{d^3 k_{\phys}}{(2\pi)^3}
\theta(H - \ome_{k, \phys}) ~|\ln (\ome_{k,\phys} / H)| \, ,
\label{Vphys}
\ee
and similarly for $S_R$. 
The step function $\theta$ restricts the integral to the amplified modes only.
In the continuum limit, this subtracts the vacuum contribution
of the high-frequency modes. In that case,
$\ome_{k, \phys} = k_{\phys}$, and the entropy densities are 
of order $H^3$ (in units where the speed of quasiparticles is equal to 1).

\section{Thermal interpretation} \label{sec:thermal}

The coefficients of $\ln |\eta_0|$ in the fits of Sec.~\ref{sec:ent},
for both dimensionalities,
are noteworthy. Earlier papers 
\cite{Brandenberger&al,Prokopec,Gasperini&Giovannini,Kruczenski&al}
have sought to define entropy of a scalar field in de
Sitter by means of a coarse-graining procedure, i.e., by neglecting the
off-diagonal elements of the density matrix in a preferred basis. The density matrix
in question is that of the entire system; in our case, that would involve the
full $M\times N$ lattice sites. The results of such coarse-graining depend on which
basis is chosen. One of the choices amounts to using the ideal gas
expression 
\be
S_{c.g.} = \sum_k \left[ (1 + n_k) \ln (1 + n_k) - n_k \ln n_k \right] ,
\label{Scg}
\ee
where $k$ goes over the normal modes of the full system (as before, we exclude 
the constant), $n_k = \sinh^2 r_k$, and $r_k$ is the squeezing parameter. According to
(\ref{Scg}), a mode with a large $|r_k|$ contributes $s_k \approx \ln n_k$ of
entropy to the total. 

We list here expressions for $n_k$ for a massless minimally coupled
scalar in two and three spatial dimensions.
In 3d, $n_k = 1/ (2 \omega_k \eta)^2$ \cite{Polarski&Starobinsky}, 
so in the limit of large squeezing $s_k \approx - 2 \ln(\ome_k |\eta|)$.
In two dimensions, the general expression for $n_k$ is
more complicated, but at large $r_k$ it simplifies into
$n_k \approx 1/|2\pi \ome_k \eta|$, so 
$s_k \approx - \ln (\ome_k |\eta|)$. (The coefficient of the logarithm
is half of that in 3d.) 
These results are obtained directly in de Sitter and make no reference to the time 
$\eta_0$ at which it ends. If, however, we substitute $\eta = \eta_0$, they
match \eqs{fitE_3d}{fitE}.\footnote{Curiously, the expression
for $n_k$ in two dimensions,
when written in terms of the physical frequency (\ref{ome_phys}) as
$n_k \approx H / (2\pi \ome_{k,\phys})$, coincides with the low-frequency tail of the Planck
distribution with the Gibbons-Hawking \cite{Gibbons&Hawking} temperature $T_{GH}= H/2\pi$.}

We find this correspondence nontrivial. For one thing, our computations do not involve
any coarse graining. For another, \eq{Scg} refers to the normal modes of the full 
system, while our results to a subsystem. Finally,
our expressions (\ref{fitE}) and (\ref{fitE_3d}) are 
for the saturation 
entropy, which obtains after both the de Sitter stage and
the subsequent growth of the EE have already ended. 
During the de Sitter stage itself, the EE is
much smaller (cf. Fig.~\ref{fig:tdep}). 
Indeed, at that time, it does not even scale as the volume of the region, 
only as the surface area
\cite{Maldacena&Pimentel}. If we think of entropy as representing decoherence (i.e., 
the loss of information on the relative phases of the basis states), 
we have to conclude that fluctuations of the field 
do not decohere significantly while they are
outside the horizon but only do so after they start oscillating. 
This, in particular, may help resolve the tension \cite{Kiefer&al} between
the large value of $S_{c.g.}$ and the amount of coherence between the first and second 
horizon crossings that is required to explain the observational data.

The occupation number $n_k$ at the end of the de Sitter stage is a definite function
of the mode frequency $\ome_k$ and
the parameter $\eta_0$. If the subsystem were isolated, the occupation numbers 
$\tn_\ell$ of its normal modes would be given by the same function,
but now of the subsystem's own normal
frequencies $\tome_\ell$. For a more 
detailed characterization of the final state, we compare the
spectrum of the parameter $f_\ell$ obtained from these $\tn_\ell$ via \eq{tnell}
to that of the thermal parameter $\beta_\ell$, \eq{beta}. We concentrate
on the limit of large squeezing, when $f_\ell \approx 1 / \tn_\ell$.
If $\tn_\ell$ obeyed the Planck distribution at temperature $T$, we would have
$f_{\ell} = \tome_\ell / T$. As it is, according to the expressions listed
after \eq{Scg},
$f_\ell \approx 2\pi \tome_\ell |\eta_0|$ in two dimensions and 
$f_\ell \approx 4 \tome_\ell^2 \eta_0^2$ in three.

\begin{figure}
\begin{center}
\includegraphics[width=4in]{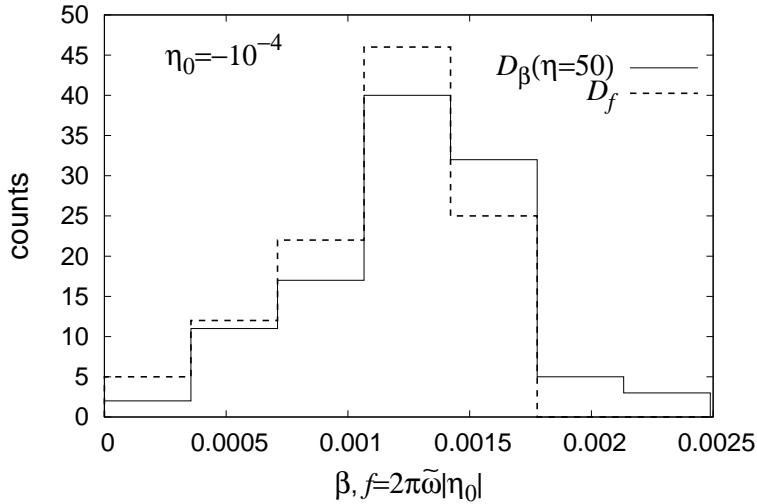}
\end{center}                                              
\caption{\small Binned distributions of the thermal
parameter $\beta_\ell$, corresponding to the Williamson normal modes of a 
subsystem at time $\eta = 50$, and 
the parameter $f_\ell$ computed from the subsystem's normal
frequencies $\tome_\ell$ as shown; $\eta_0 = - 10^{-4}$.
The results are for a massless scalar in two spatial dimensions, for a subsystem defined as a
$10\times 11$ rectangle in the middle of a  
$31\times 33$ square lattice.
}
\label{fig:hist}  
\end{figure}

Even if were assured of thermalization, we could not
expect a perfect correspondence between the individual $\beta_\ell$ and $f_\ell$.
Indeed, $f_\ell$ are computed under the assumption that the subsystem is perfectly isolated,
while in practice it is not. So, for a comparison, 
we bin $\beta_\ell$ and $f_\ell$ into distributions
$D_\beta(\eta)$ and $D_f$, which
count the number of times the 
corresponding parameter ($\beta_\ell$ or $f_\ell$)
falls into a bin centered on a given value. Note that $D_\beta$
depends on time, while $D_f$, being fully determined by $\eta_0$ and the spectrum of
the normal modes, does not.
A sample result for two dimensions is shown in Fig.~\ref{fig:hist}.
Results for three dimensions are similar. We wish to stress that a good agreement between
the two distributions, seen in Fig.~\ref{fig:hist}, develops only at sufficiently large times.
At $\eta =0$, $\beta_\ell$ are distributed over a much broader range.
As discussed in the Introduction, we interpret the agreement at large times
as a restricted version of thermalization.

\section{Conclusion}
Our main conclusion is that a thermal interpretation 
(along the lines of Sec.~\ref{sec:thermal}) of the entanglement 
entropy of a Gaussian scalar in a region of the de Sitter spacetime 
looks plausible, provided one applies it not to fluctuations
in that spacetime
itself but rather to those that evolve after the de Sitter stage has ended. 
It is as if the spacetime stores and then releases latent heat. If experimental 
results for the quantum simulator (the Bose-Hubbard model) described in Sec.~\ref{sec:lat}
become available, comparing them to the results obtained here for the Gaussian theory
may elucidate the role of nonlinearities.

\section*{Acknowledgments}
We thank Chen-Lung Hung and 
Martin Kruczenski for discussions. This work was supported in part by the U.S.
Department of Energy (grant DE-SC0019202) and by the W. M. Keck Foundation.

\end{document}